\documentclass[reqno,11pt]{article}  

\newlength{\dinwidth}
\newlength{\dinmargin}
\usepackage{amsmath}
\usepackage{amssymb}
\usepackage{latexsym}
\usepackage{cite}
\usepackage{epsfig,psfrag}
\newcommand{\RR}{\mathbb{R}}

\newcommand{\CC}{\mathbb{C}}

\newcommand{\pihalf}{{\frac{\pi}{2}}} 
 
\newcommand{\unity}{{\setlength{\unitlength}{1em}
                     \begin{picture}(0.75,1)
                     \put(0,0){$1$}
                     \put(0.34,0){\line(0,1){0.65}}
                     \end{picture}
                   }}
\newcommand{\unit}{{\mbox{\texttt 1}}} 
\renewcommand{\Im}{\text{Im}\,}

\newenvironment{Proof}%
{\par \medskip \noindent {\em Proof.}}{\hspace*{\fill} $\square$\par%
\medskip\noindent}
\newtheorem{Thm}{Theorem}[section]
\newtheorem{Prop}{Proposition}
\newtheorem{Lem}{Lemma}
\newtheorem{Cor}[Lem]{Corollary}

\newcommand{\charge}{\chi} 
\newcommand{\concharge}{{\bar{\chi}}} 
\newcommand{\calH}{{\mathcal H}}
\newcommand{\Heins}{\calH_{\charge}^{(1)}} 
\newcommand{\HeinsConj}{\calH_{\concharge}^{(1)}} 
\newcommand{\Eeins}{E_\charge^{(1)}}     
\newcommand{\EeinsConj}{E_{\concharge}^{(1)}} 
\newcommand{\adj}{\dagger}       
\newcommand{\calA}{\mathcal{A}}          
\newcommand{\calF}{\mathcal{F}}          
\newcommand{\calO}{\mathcal{O}}          
\newcommand{\calK}{\mathcal{K}}          
\newcommand{\calU}{\mathcal{U}}          
\newcommand{\calUtild}{\tilde{\mathcal{U}}}          
\newcommand{\strip}{G}    
\newcommand{\Potild}{\tilde{P}_+^{\uparrow}}
\newcommand{\Po}{P_+^{\uparrow}}
\newcommand{\Lortild}{\tilde L_+^\uparrow}
\newcommand{\Lor}{L_+^{\uparrow}}
\newcommand{\half}{{\frac{1}{2}}} 
\newcommand{\act}{\!\cdot\!}
\newcommand{\clo}{ {\mbox{\bf --}} }
\newcommand{\Wtild}{{\tilde{W}}}
\newcommand{\Hyp}{{H_m^+}}

\newcommand{\HypC}{{H_m^c}}

\newcommand{\eps}{\varepsilon}
\newcommand{\alp}{\alpha}
\newcommand{\bet}{\beta}
\newcommand{\gam}{\gamma}
\newcommand{\We}{\Wtild_1}             
\newcommand{\spc}{{C}}             
\newcommand{\spcpath}{\tilde{\spc}}    
 
\newcommand{\spcR}{\boldsymbol{\spc}} 
\newcommand{\xR}{\boldsymbol{x}}          
\newcommand{\yR}{\boldsymbol{y}}          
\newcommand{\pR}{\boldsymbol{p}}          
\newcommand{\kR}{\boldsymbol{k}}          
\newcommand{\spd}{e}                   
\newcommand{\Spd}{H}              
\newcommand{\spdpath}{\tilde{e}}       
\newcommand{\refspd}{{e_0}}            
\newcommand{\refspdpath}{\spdpath_0}  
\renewcommand{\lor}{\lambda}     
\newcommand{\lortild}{\tilde{\lor}} 
\newcommand{\po}{g}              
\newcommand{\potild}{\tilde{\po}}  
\newcommand{\rottild}[1]{\tilde{r}(#1)}  
\newcommand{\rot}[1]{r(#1)}          
\newcommand{\boox}[1]{\tilde{\lambda}_1(#1)}   
\newcommand{\Boox}[1]{\lambda_1(#1)}           
\newcommand{\J}{j}         
\newcommand{\Auni}{\calA_{\text{uni}}}
\newcommand{\HVac}{\calH_{00}}
\begin{document} 
\title{The Spin-Statistics Theorem for Anyons and Plektons 
in d=2+1}
\author{Jens Mund\thanks{Supported by FAPEMIG 
.}
\\ 
\scriptsize 
Departamento de F\'{\i}sica, Universidade Federal de Juiz de Fora,\\  
\scriptsize 
36036-900 Juiz de Fora, MG, Brazil.\\ 
\scriptsize  E-mail: {\tt mund@fisica.ufjf.br}}
\date{
{\em 
Dedicated to Klaus Fredenhagen on the occasion of his 
60th birthday.}
}
\maketitle 
\begin{abstract}
We prove the spin-statistics theorem for massive particles  obeying
braid group statistics in three-dimensional Minkowski space. 
We start from first principles of local relativistic quantum theory. 
The only assumption is a gap in the mass spectrum of the 
corresponding charged sector, and a restriction on the degeneracy of
the corresponding mass.
\end{abstract}
\section{Introduction} 
The famous spin-statistics theorem relates the exchange statistics of
a quantum field with the spin of its elementary excitations~\cite{SW}. Namely,
it states that in the case of Bose/Fermi (para-) statistics there holds 
$$
e^{2\pi i s} = \text{sign } \lambda,
$$
where $s$ is the spin of the particles and $\lambda$ is the
statistics parameter of the fields. 
It has found in~\cite{DHRIII} a derivation from first
principles without any non-observable quantities such as
charge-carrying fields. 
However, a basic input to this derivation was that the charge be 
localizable in bounded regions. In~\cite{BuEp}, Buchholz and Epstein
extended the theorem to massive particles carrying a non-localizable
charge. In the purely massive case, such charges are still localizable 
in space-like cones~\cite{BuF}, i.e., cones in spacetime which extend
to space-like 
infinity.\footnote{More precisely, a space-like cone is a region in 
  Minkowski space of the form $\spc=a+\cup_{\lambda>0}\lambda \calO,$ where
  $a\in\RR^d$ is the apex of $\spc$ and $\calO$ is a double cone whose
  closure does not contain the origin.} 
The analysis of Buchholz and Epstein was carried out in
four-dimensional spacetime, in which
case $\lambda$ is a real number associated with a unitary representation of
the permutation group ($\lambda>0$ corresponding to Bosons and
$\lambda<0$ corresponding to Fermions). 
In three-dimensional spacetime, however, it may occur that the
permutation group is replaced by the braid group, in which case the
statistics parameter is a complex non-real number. 
The phase in its polar decomposition is called the statistics phase
$\omega$, 
\begin{equation}\label{eqStatPhase}
\omega := \frac{\lambda}{|\lambda|}. 
\end{equation} 
In the case of non-real $\lambda$ (i.e.\ $\omega\neq\pm1$) one speaks of
braid group statistics 
and calls the particles Plektons or, if the corresponding representation is 
Abelian, Anyons. 
Related to this phenomenon, in 
three-dimensional spacetime the spin of a particle needs not be
integer or half-integer, but may assume any real value (``fractional'' spin). 
In fact, the occurrence of braid group statistics is equivalent 
to the occurrence of ``fractional'' spin.

In the present article, we prove that in this case the spin-statistics relation
\begin{equation}\label{eqSpiSta}
e^{2 \pi i s} = \omega 
\end{equation}
holds,   starting from first principles and only assuming the following 
conditions on the mass spectrum. 
We consider a charged sector of a local relativistic quantum theory  
in three-dimensional Minkowski space, containing a massive particle
with mass $m>0$ and spin $s\in\RR$. We assume that $m$ is separated from
the rest of the mass spectrum in its sector by a mass gap.  
We further assume that
there are only finitely many ``particle types'' in its sector with
this mass, and that they all have the same spin $s$. 
As a byproduct, we prove that the familiar symmetry between particles
and antiparticles holds also in this case: 
Namely, that there is an equal number of antiparticle types 
(in the conjugate sector) with the same
mass which all have the same spin $s\in\RR$ (Proposition~\ref{AntiPart}). 

It should be noted that a ``weak spin-statistics  
relation'',
\begin{equation} \label{eqSpiStaWeak}
e^{4 \pi i s} = \omega^2, 
\end{equation}
is known to hold~\cite{FredSum,FG} under quite general conditions
in the case of braid group statistics. 
It should also be noted that the strong spin-statistics
relation~\eqref{eqSpiSta} has been proved in~\cite{FM2} and in~\cite{Longo96},
but under a non-trivial hypothesis ammounting to the
Bisognano-Wichmann property, or modular covariance, of the
charged fields~\cite{FM2} or the observables~\cite{Longo96}, respectively. 
In the present
paper we do not need this hypothesis. In fact, we shall show in a 
subsequent paper~\cite{Mu_BiWiAny} that the Bisognano-Wichmann property
may be derived from first principles in a purely massive theory with
braid group statistics, using the results of our present analyisis. 

Our derivation will largely parallel that of Buchholz and
Epstein~\cite{BuEp}.    
The crucial difference between the four-dimensional case considered in
\cite{BuEp} and the present three-dimensional case lies in the
structure of the Poincar\'e group and the irreducible massive
representations of its universal covering group, which have been
heavily used in~\cite{BuEp}. In particular, in four dimensions one has
the so-called ``covariant representation''\footnote{This is a tensor
  product of the spin zero representation of the Poincar\'e group with a
  finite-dimensional representation of the (covering of the) Lorentz
  group.}, in which locally
generated single particle wave functions have certain analyticity 
properties which are exploited in the proof. In three dimensions,
however, there is no ``covariant represention'', and in the well-known
Wigner representation the wave functions are {\em not} analytic. As a way
out, we use here an equivalent representation found by the author in
\cite{M02a}, which exhibits precisely the required analyticity
properties. On the other hand, the representation of the translation
subgroup in three dimensions does not differ essentially from that in
four dimensions. Hence the results from~\cite{BuEp} which use only
the translations can directly be adapted to the three-dimensional
case. This concerns in particular our Lemma~\ref{BuEp}.

The article is organized as follows. In Section~\ref{SecAss} we specify
in detail our framework, assumptions and results.  
In Section~\ref{Sec2PtPAP} we recall a result of Buchholz and
Epstein~\cite{BuEp} 
concerning analyticity of the two-point functions in momentum space,
and extend their result on the particle-antiparticle symmetry to the
present case. 
In Section~\ref{SecSpiSta}, finally, we prove the spin-statistics theorem. 
\section{Framework, Assumptions and Results} \label{SecAss}
We now specify our framework and make our assumptions and results 
precise.\footnote{
Recall that in the case of braid group statistics there
is no canonical way to construct a field algebra from
the observables~\cite{MuRe}. But our framework, using a
restricted notion of charged fields, can be 
set up starting from the standard assumptions~\cite{H96} of local relativistic
quantum theory on the observabels plus weak Haag
  Duality, together with our assumptions on the mass spectrum. 
{}For the convenience of the reader, 
we sketch in Appendix~\ref{Just} how this may be done and 
indicate the relation 
with the notions used in the literature~\cite{BuF,DHRIV,FGR}.}
\paragraph{States and Fields.}
Denoting the quantum numbers of our sector collectively by $\charge$,  
the space of {states} of the sector corresponds to a 
Hilbert space $\calH_\charge$. 
It is orthogonal to the the vacuum Hilbert space $\calH_0$ which 
contains a Poincar\'e invariant vector $\Omega$, corresponding to 
the vacuum state.  
$\calH_\charge$ carries a unitary representation of the universal 
covering group $\Potild$ of the Poincar\'e group  in 2+1 dimensions, 
denoted by $U_\charge$, satisfying the relativistic spectrum condition
(positivity of the energy). 
{Fields} carrying charge $\chi$ are bounded operators from
the vacuum Hilbert space $\calH_0$ to $\calH_\charge$. 
The linear space of these fields will be denoted by $\calF_\charge$.  
\paragraph{Localization.} 
Fields are {localizable} to the same extent to which the charges are
localizable which they carry. 
In the case of braid group statistics, the charges cannot be localized in
bounded regions of spacetime~\cite{DHRIII}, but they can be localized,
in the massive case,  
in regions which extend to infinity in some space-like direction, 
namely, in space-like cones~\cite{BuF}.
Now the manifold of space-like directions, 
\begin{equation} \label{eqSpd} 
\Spd :=\{ e\in\RR^3,\; e\cdot e=-1\}, 
\end{equation} 
is {not} simply connected in three dimensions (in contrast to the
four-dimensional case): Given two space-like directions, there exists
an infinity of non-homotopic paths in $\Spd$ from one to the other,
distinguished by a winding number. It is precisely this fact which
enables the occurrence of braid group statistics in three dimensions
(see the remark after Eq.~\eqref{eqCommut} below). 
To realize such statistics, the {fields} which create a charge
localized in a given space-like cone $\spc$ need an additional
information: Namely, a path in the set of space-like
directions $\Spd$ starting from some fixed reference 
direction $\refspd$ and ``ending'' in $\spc$.\footnote{Two other 
possibilities are:
To introduce a reference space-like cone from
which all allowed localization cones have to keep space-like separated 
(this cone playing the role of a ``cut'' in the
context of multivalued functions)~\cite{BuF}; or a cohomology theory of 
nets of operator algebras as introduced by 
Roberts~\cite{Roberts,Roberts76,Roberts80}.} 
We shall sketch this concept, which has been introduced
in~\cite{FRSII}, in a slightly modified form introduced in~\cite{MDiss}. 
We say that a space-like cone $\spc$ {contains a 
space-like direction} $\spd$ if 
\begin{equation} \label{eqDirInSpc}
 \spc+e\subset \spc \,.
\end{equation}
We say that a path $\spdpath$ in $\Spd$ {\em ends in}
$\spc$ if its endpoint is contained in $\spc$ in the sense of
equation~\eqref{eqDirInSpc}.  
Two paths $\spdpath_1$ and $\spdpath_2$
starting at $\refspd$ and ending in $\spc$ will be called 
{equivalent w.r.t.\ }$\spc$ iff the path $\spdpath_2\ast\spdpath_1^{-1}$ 
(the inverse of $\spdpath_1$ followed by $\spdpath_2$) is
fixed-endpoint homotopic to a path which is contained in $\spc$.  
Figure~1 illustrates this concept. 
By a {\em path of space-like cones} we shall understand 
a pair 
\begin{equation} \label{eqSpcpath} 
(\spc,\spdpath)  \,, 
\end{equation}
where $\spc$ is a space-like cone and $\spdpath$ is the equivalence
class w.r.t.\ $\spc$ of a path in $\Spd$ starting at $\refspd$ and 
ending in $\spc$. 
(We use the same symbol for a path and its equivalence class.) 
{}We shall use the notation $\spcpath$ for a
path of space-like cones of the form $(\spc,\spdpath)$. 
%
Such paths of space-like cones serve to label the localization
regions of charged fields. 
Namely, for each $\spcpath$ there is a linear subspace 
$\calF_\charge(\spcpath)$ of $\calF_\charge$, called the fields carrying
charge $\charge$ localized in $\spcpath$. 
This family is isotonuous in the sense that 
\begin{equation} \label{eqIsotony} 
\calF_\charge(\spcpath_1)\subset \calF_\charge(\spcpath_2) \qquad 
\text{ if } \qquad \spcpath_1\subset \spcpath_2. 
\end{equation}
(We say that $\spcpath_1\doteq(\spc_1,\spdpath_1)$ is contained in 
$\spcpath_2\doteq(\spc_2,\spdpath_2)$, in symbols  
\begin{equation} \label{eqSpcSub}
\spcpath_1 \subset \spcpath_2 \,, 
\end{equation}
if $\spc_1\subset \spc_2$  and
the corresponding paths $\spdpath_1$, $\spdpath_2$ are equivalent 
w.r.t.\ $\spc_2$.) 
The vacuum $\Omega$ has the Reeh-Schlieder property for the fields,
i.e.\ for any path of space-like cones $\spcpath$ holds 
\begin{equation} \label{eqReehSchlie}
\big(\calF_\charge(\spcpath)\,\Omega\big)^\clo = \calH_\charge, 
\end{equation}
where the bar denotes the closure. 
\begin{figure}[ht] 
 \label{Fig1}
\psfrag{e0}{$e_0$}
\psfrag{e1}{$\spdpath_1$}
\psfrag{e2}{$\spdpath_2$}
\psfrag{e3}{$\spdpath_3$}
\psfrag{C}{$\hat C$}
\psfrag{H}{$H$}
\begin{center}
\epsfxsize35ex 
\epsfbox{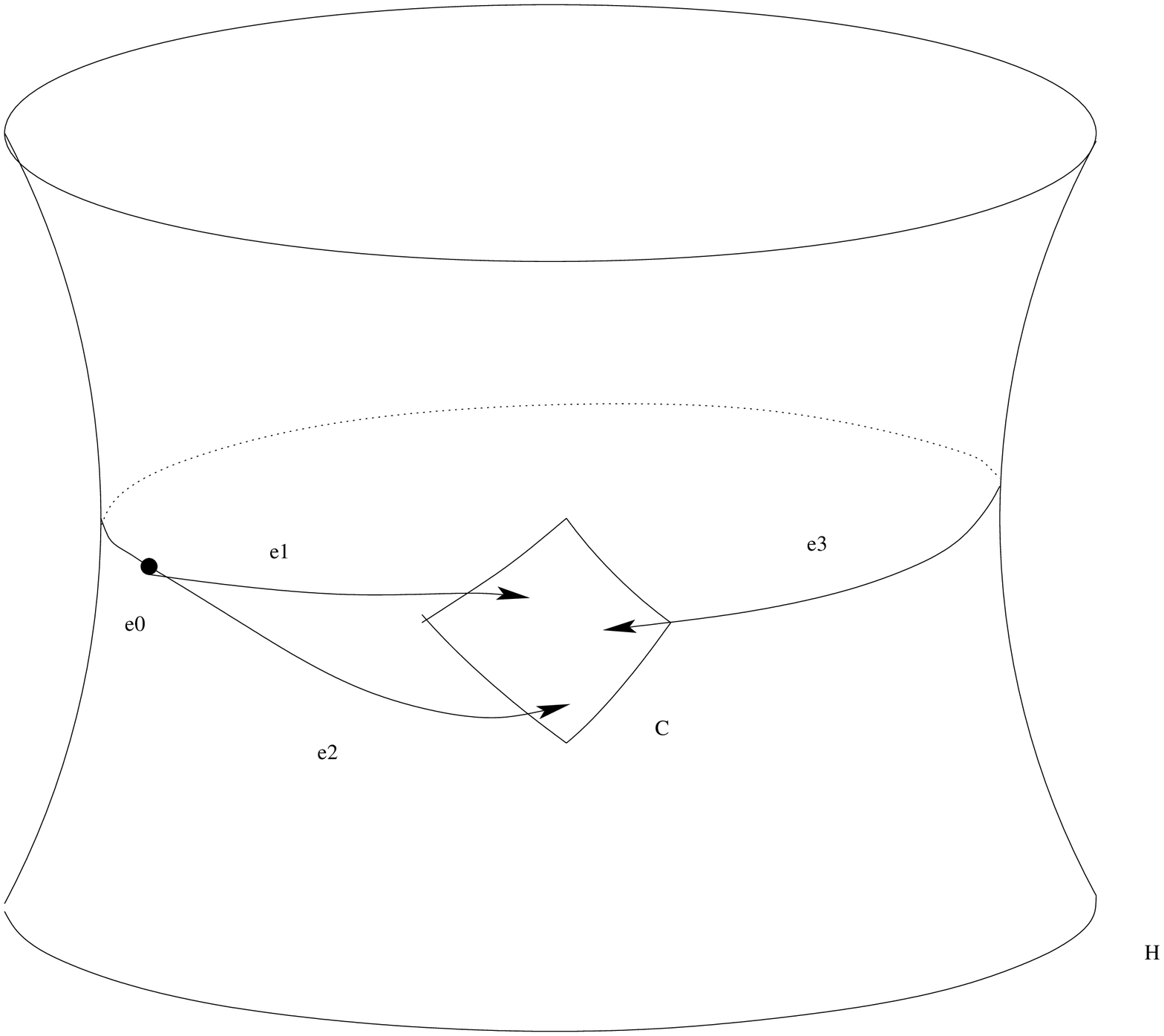}
\caption{$\hat C$ denotes the set of space-like directions contained in
  $C$, in the sense of Eq.~\eqref{eqDirInSpc}. $(C,\spdpath_1)$ is 
equivalent with $(C,\spdpath_2)$, but
  inequivalent from $(C,\spdpath_3)$.}
\end{center}
\end{figure}
\paragraph{Covariance.} 
There is a representation $\alpha_\charge$ 
of the universal covering group $\Potild$ of the Poincar\'e group
$\Po$  by endomorphisms of $\calF_\charge$, which implements the
unitary representation $U_\charge$ in the sense that 
\begin{equation} \label{eqUImplement} 
\alpha_\charge(\potild)(F)\,\Omega = U_\charge(\potild)\,F\,\Omega 
\end{equation}
holds for all $\potild\in\Potild$ and $F\in\calF_\charge$. 
It acts covariantly on the fields in the following sense: 
\begin{equation} \label{eqCov}
\alpha_\charge(\potild):\calF_\charge(\spcpath)\to
\calF_\charge(\potild\act\spcpath). 
\end{equation}
Here, $\potild\cdot\spcpath$ denotes the natural action of the 
universal covering of the Poincar\'e group on the paths of
space-like cones, defined as follows. Let $\potild=(a,\lortild)$, where 
$a$ is a spacetime translation and $\lortild$ is an element of the
universal covering group $\Lortild$ of the Lorentz group, projecting onto
$\lor\in\Lor$. Then 
\begin{equation} \label{eqPoincSpc}
\potild\act (\spc,\spdpath):= (g\act \spc,\lortild\act \spdpath),
\end{equation}
where $\lortild\act \spdpath$ denotes the lift of the action of 
the Lorentz group on $\Spd$ to the respective universal covering spaces. 
Note that a $2\pi$ rotation acts non-trivial --- 
it maps, for example, $(C,\spdpath_3)$ in Figure~1 onto $(C,\spdpath_1)$.  
\paragraph{Conjugate Charge.} 
There is a sector with the conjugate charge $\concharge$, for which
all of the above-mentioned facts also hold. 
We shall denote the corresponding objects by 
$\calH_\concharge$, $U_\concharge$, $\calF_\concharge(\spcpath)$, and
$\alpha_\concharge$, respectively. In particular,
$\calF_\concharge(\spcpath)$ is a linear space of operators mapping
$\calH_0$ onto $\calH_\concharge$.  
There is a notion of operator adjoint, 
which associates with each field $F\in \calF_\charge$ an adjoint field 
operator  $F^\adj\in\calF_\concharge$, satisfying
$(F^\adj)^\adj=F$ and preserving localization, i.e.\ 
\begin{equation}\label{eqFadjLoc}  
\big(\calF_\charge(\spcpath)\big)^\adj = \calF_\concharge(\spcpath).
\end{equation}
The operation of adjoining intertwines the representations $\alpha_\charge$ and
$\alpha_\concharge$ in the sense that 
\begin{equation} \label{eqalphaConj}
\big(\alpha_\charge(\potild)(F)\big)^\adj =
\alpha_\concharge(\potild)(F^\adj).
\end{equation}
\paragraph{Statistics.} 
There is a complex number $\omega_\charge$ of modulus one, the
statistics phase  
of the sector $\charge$, which (partly) characterizes the statistics
of fields. Namely, suppose $\spcpath_1=(\spc_1,\spdpath_1)$ and 
$\spcpath_2=(\spc_2,\spdpath_2)$ are such that $\spc_1$ and $\spc_2$ are
causally separated, and the path $\spdpath_1 \ast \spdpath_2^{-1}$ 
goes ``directly'' from $\spc_2$ to $\spc_1$ in the mathematically
positive sense.\footnote{``Directly'' means that it stays 
causally separated from the cone $\spc_2$ once it has left it; and
``mathematically positive sense'' means here the right-handed
sense w.r.t.\ a future pointing time-like Minkowski vector.} 
(Note that this condition is independent of the choice of 
reference direction $\spd_0$. 
Figure~2 shows an example satisfying these conditions.)  
Then for $F_i\in\calF_\charge(\spcpath_i)$, $i=1,2$, there holds 
\begin{equation}\label{eqCommut}
\big(\, F_2\Omega,F_1\Omega\,\big) = 
\omega_\charge \,\big(\, F_1^\adj\Omega,F_2^\adj\Omega\,\big).
\end{equation}
Note that the hypothesis  under which
Eq.~\eqref{eqCommut} holds is not symmetric in $\spcpath_1$ and
$\spcpath_2$ just because of the condition on the paths $\spdpath_i$. 
Without this condition, Eq.~\eqref{eqCommut} would imply
$\omega_\charge\omega_\concharge=1$. But $\omega_\charge$ and
$\omega_\concharge$ are known to coincide~\cite{FM2}, hence 
Eq.~\eqref{eqCommut} would be be self-consistent only
for $\omega_\charge=\pm 1$, excluding braid group statistics.  
\begin{figure}[ht] 
 \label{Fig2}
\psfrag{e0}{$e_0$}
\psfrag{e1}{$\spdpath_1$}
\psfrag{e2}{$\spdpath_2$}
\psfrag{C1}{$\hat{C}_1$}
\psfrag{C2}{$\hat{C}_2$}
\psfrag{H}{$H$}
\begin{center}
\epsfxsize35ex 
\epsfbox{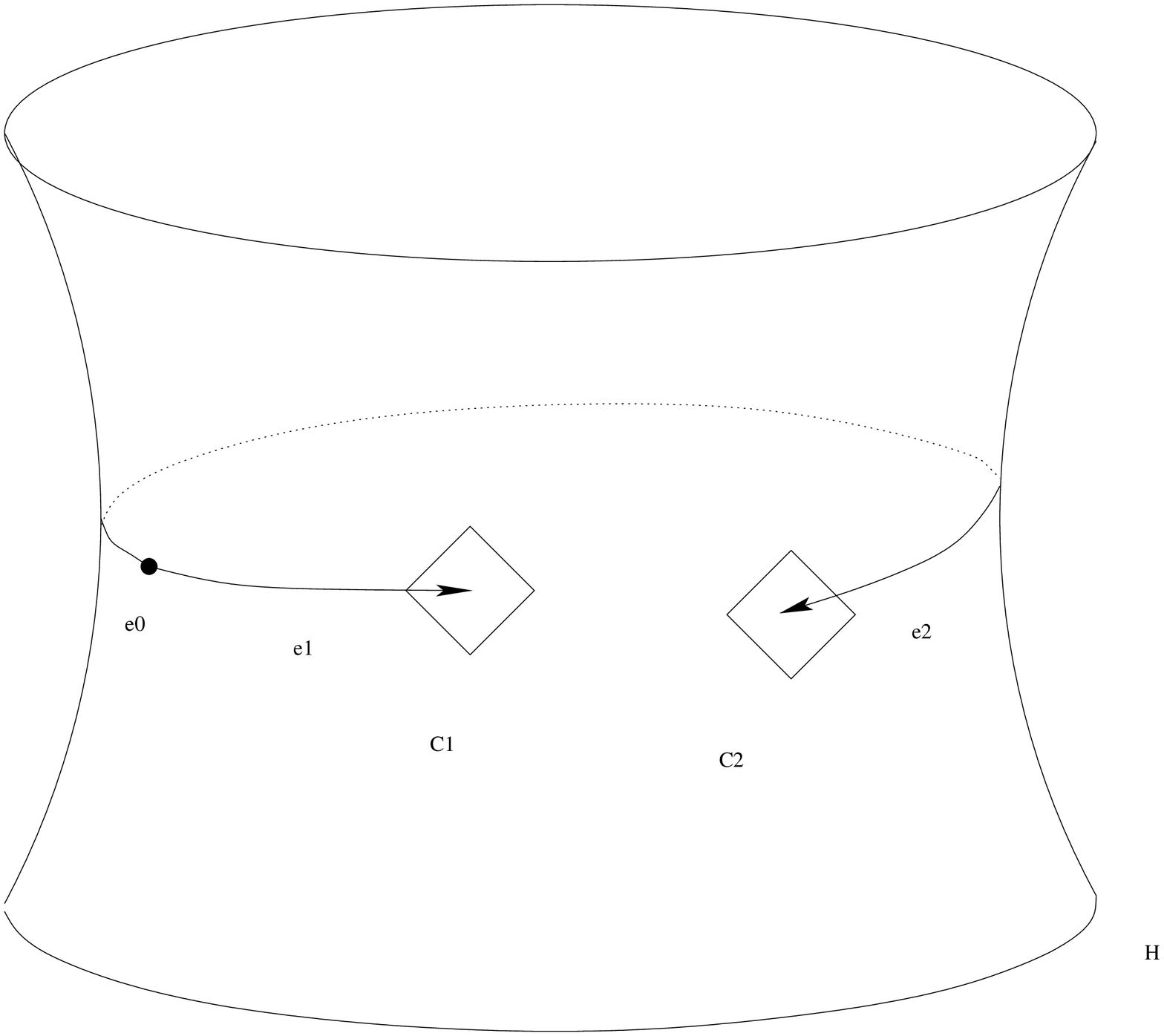}
\caption{$(C_1,\spdpath_1)$ and $(C_2,\spdpath_2)$ satisfy the
  hypothesis under which Eq.~\eqref{eqCommut} holds.} 
\end{center}
\end{figure}
\paragraph{Assumptions on the Particle Spectrum.} 
We consider a particle of stricly
positive mass $m$ and spin $s$ 
in the sector $\charge$, and assume that $\{m\}$ is separated from 
the rest of the mass spectrum in the sector $\charge$ by a mass gap. 
We further assume that
there are only finitely many ``particle types'' in the sector  
$\charge$ with this mass, and that they all have the same spin $s$. 
More technically, let $P_\charge$ be the 
energy-momentum operator in the sector $\charge$, i.e.\ the vector
operator which generates the spacetime translations in the sense that 
$U_\charge(a)=\exp (ia\cdot P_\charge)$ for $a\in\RR^3$, and let
$M_\charge:=P_\charge^2$ be the mass operator in the sector $\charge$.  
This operator has as an eigenvalue the mass, $m$, of our
particle. Our {assumptions} then are: 
\begin{itemize} 
\item[{(A1)}] The mass $m$ is strictly positive.\label{eqPos} 
\item[{(A2)}] $m$ is an isolated point in the spectrum of $M_\charge$.
  \label{eqIso}   
\item[{(A3)}] 
The restriction of the representation $U_\charge$ to the corresponding 
eigenspace is \label{eqFinite} 
a finite multiple of the irreducible representation with 
mass $m$ and spin $s$.
\end{itemize}
It is gratifying that the assumptions (A1) and (A2), together with
the standard assumptions on the observables plus weak duality, 
imply the validity of our entire framework. In particular, they imply that
the charge $\charge$ is localizable in space-like cones~\cite{BuF} and 
allow for the determination of the statistics phase 
$\omega_\charge$ (namely, they exclude the so-called infinite statistics, 
$\lambda=0$~\cite{F81}). 
\paragraph{Results.}
Under the above assumptions (A1) through (A3), we shall prove that the strong
spin-statistics relation~\eqref{eqSpiSta} holds 
in the case of braid group statistics (Theorem~\ref{SpiSta}). As a  
byproduct, we prove that the familiar
symmetry between particles and antiparticles holds also in this case. 
Namely, it is 
known that the mass spectrum of the conjugate sector $\concharge$ 
coincides with that of $\charge$~\cite{F81}, and that the spins
occuring in the eigenspace corresponding to mass $m$ in the sector $\charge$ 
coincide with those in the conjugate sector $\concharge$ modulo 
one~\cite{FM2}. What we show is that the spins actually coincide as
real numbers, and that the 
degeneracies in the conjugate sectors $\charge,\concharge$ coincide
--- in other words, that the corresponding ray representations of the 
Poincar\'e group are unitarily equivalent (Proposition~\ref{AntiPart}). 
\section{Momentum Space Two-Point Functions and
  Particle-Antiparticle Symmetry} \label{Sec2PtPAP}
Buchholz and Epstein's proof of the spin-statistics theorem in four
dimensions relies on their result on the two-point
functions in momentum space~\cite{BuEp}. The latter result extends 
straightforwardly to the present three-dimensional case, because it has been
derived under precisely our conditions of covariance~\eqref{eqCov}, 
the Reeh-Schlieder property~\eqref{eqReehSchlieEins}, commutation
relations as in Eq.~\eqref{eqCommut} and a mass gap around $m>0$,
without referring to the representation of the Lorentz subgroup 
(which makes the crucial difference between three and
four dimensions). 

To state their result, some notation needs to be introduced. 
Fixing a Lorentz frame, spacetime points are written
as $x=(x^0,\xR)$, 
and the Minkowski scalar product reads 
$(x^0,\xR)\cdot(y^0,\yR)=x^0y^0-\xR\cdot\yR$, where $\xR\cdot\yR$ 
denotes the standard scalar product in $\RR^2$.
The positive and negative mass shells $H_m^{\pm}$ are 
the set of momentum space points $p=(p_0,\pR)$ satisfying 
$p_0^2-\pR\cdot \pR=m^2$ and $p_0\gtrless 0$, respectively.   
The unique (up to a factor) Lorentz invariant measure on $\Hyp$ 
is denoted by $d\mu(p)$.   
The complexified mass shell $\HypC$ is defined as the set of
$k=(k_0,k_1,k_2)$ $\in\CC^3$ satisfying $k_0^2-k_1^2-k_2^2=m^2$. 
Buchholz and Epstein consider a special class of space-like
cones, namely, those of the form 
\begin{equation} \label{eqSpcR}
\spc = \spcR'',
\end{equation}
where $\spcR$ is an open, salient cone with apex at the origin 
in the rest space 
of the fixed frame (which we shall occasionally identify with $\RR^2$), 
and $\spcR''$ denotes its causal completion.
For a cone $\spcR$ of this form, let its dual $\spcR^*$ be defined by 
\begin{equation} \label{eqDualCone}
\spcR^* :=\big\{ \pR\in \RR^2: \pR\cdot \xR>0 \; \forall \xR\in
\spcR^\clo\setminus\{0\} \,\big\}.
\end{equation}
Buchholz and Epstein use regularized fields, for which the functions
$\potild\mapsto \alpha_\charge(\potild)(F)$ are smooth. The set of
smooth fields carrying charge $\charge$ and localized in $\spcpath$
shall be denoted by $\calF_\charge^\infty(\spcpath)$. The
Reeh-Schlieder property~\eqref{eqReehSchlie} still holds for the
smooth fields, also on the single particle space. More precisely, 
let $\Eeins$ be the spectral projector of the mass operator 
corresponding to the eigenvalue $m$, and let $\Heins$ be its range, i.e.\ the
corresponding eigenspace.  
Then there holds 
\begin{equation}\label{eqReehSchlieEins}
\big(\Eeins\calF_\charge^\infty(\spcpath)\,\Omega\big)^\clo = \Heins. 
\end{equation}
The result of Buchholz and Epstein on the two-point functions, in
the present context, is the following: 
\begin{Lem}[Buchholz, Epstein]\label{BuEp}
Let $\spc_1$ and $\spc_2$ be causally separated space-like cones of
the form~\eqref{eqSpcR} such that $\spcR_{12}:=\spcR_2-\spcR_1$
is a salient cone, and let $\spcpath_1,\spcpath_2$ be such that the
hypothesis of Eq.~\eqref{eqCommut} is satisfied. 
Then for any pair of fields $F_i\in\calF_\charge^\infty(\spcpath_i)$,
$i=1,2$, there exists a function $h$ which is analytic in the region 
\begin{equation} \label{eqDomain} 
\Gamma:=\{k=(k_0,\kR)\in\HypC:\, 
\Im\kR\in\big(\spcR_{12}\big)^*\,\}  
\end{equation}
and has smooth boundary values on the mass shells $H_m^{\pm}$
satisfying 
\begin{align} \label{eq2PtFct}
\big(\, F_2\Omega,U_\charge(x) \Eeins F_1\Omega\,\big) & = 
\int_\Hyp d\mu(p)\, h(p)\,e^{ip\cdot x}, \\
\omega_\charge \,\big(\, F_1^\adj\Omega,
U_\concharge(x)\EeinsConj F_2^\adj\Omega\,\big) &= 
\int_\Hyp d\mu(p)\,h(-p)\,e^{ip\cdot x}.
\end{align}
\end{Lem}
\begin{Proof}
Replacing the factor ``sign $\lambda$'' in Eq.~(2.2) of~\cite{BuEp} by 
our $\omega_\charge$, Buchholz and Epstein's proof 
can be directly transferred to the present setting, since it uses only 
the conditions of covariance~\eqref{eqCov}, space-like commutation 
relations~\eqref{eqCommut}, Reeh-Schlieder
property~\eqref{eqReehSchlieEins} and a mass gap around $m>0$.
\end{Proof}
The Lemma immediately implies the existence of antiparticles
with the same mass $m$ as the particles in the sector $\charge$ (which
had been established in this generality already in~\cite{F81}).  
Moreover, it implies a complete symmetry between particles and
antiparticles, valid also in the present case of braid group
statistics in three dimensions: 
\begin{Prop}[Particle-Antiparticle Symmetry] \label{AntiPart} 
The spins and multiplicities of the single particle spaces $\Heins$
and $\HeinsConj$ coincide. In particular, the restriction to $\HeinsConj$ of
the representation $U_\concharge$ is equivalent with the restriction to
$\Heins$ of the representation $U_\charge$. 
\end{Prop}
\begin{Proof}
The proof requires only a slight modification from that of Buchholz
and Epstein. Namely, the role of the square of the Pauli-Lubanski
vector as a Casimir operator is, in 2+1 dimensions, played by a 
scalar operator, the so-called Pauli-Lubanski
scalar~\cite{Binegar,JackiwNair} which is defined as follows. Let $U$ be a
representation of the universal covering of the Poincar\'e group in
three spacetime dimmensions, let $L_0$ denote the generator of the  
rotation subgroup in the representation $U$, and let $L_i$ be the
generator of the boosts in direction $x^i$, $i=1,2$. Let further
$J_\mu$ be the vector operator $J_\mu=(-L_0,L_2,-L_1)$. 
The Pauli-Lubanski scalar of the representation $U$ is defined as 
\begin{equation}\label{eqPauliLub}
W:= J_\mu P^\mu, 
\end{equation}
where $P^\mu$ are the generators of the translation subgroup in the
representation $U$. It has the following
properties~\cite{Binegar,JackiwNair}: 
it commutes with the representation $U$, and has the value 
\begin{equation}\label{eqPauliLub1} 
W = -ms\unity 
\end{equation}
if, and only if,  $U$ contains only irreducible representations whose
masses and spins have the product value $ms$. 
Considering now the representations $U_\charge$ and $U_\concharge$, we 
denote their Pauli-Lubanski scalars as $W_\charge$ and $W_\concharge$,
respectively. The key point is that for each field 
$F\in\calF_\charge^\infty(\spcpath)$, there is a field
$\delta_\charge(F)\in\calF_\charge^\infty(\spcpath)$ such that, due to
covariance~\eqref{eqUImplement}, there holds\footnote{Namely, 
$\delta_\charge$ is the ``derivation'' on
$\calF_\charge^\infty$ defined by 
$$
\delta_\charge (F):= - \frac{d}{ds}\,\frac{d}{dt} 
\sum_{\mu=0}^2
\alpha_\charge\big(\lortild^{(\mu)}(t)\, T(se_{(\mu)})\big)
\big(F\big) \big|_{s=t=0}, 
$$
where $T(\cdot)$ is the translation subgroup, $e_{(\mu)}$ are the
unit vectors in the given Lorentz frame,  
$\lortild^{(0)}(-t)$ is the rotation subgroup, $\lortild^{(1)}(t)$
is the boost subgroup in direction $e_{(2)}$ and $\lortild^{(2)}(-t)$
is the boost subgroup in direction $e_{(1)}$.}
\begin{equation}\label{eqdeltaFOmega}
W_\charge F\Omega=\delta_\charge(F)\Omega.
\end{equation}
The same holds for the
conjugate sector $\concharge$. Let now, for $i=1,2$, $\spcpath_i$ and 
$F_i\in\calF_\charge^\infty(\spcpath_i)$ satisfy the hypothesis of 
Lemma~\ref{BuEp}. Then 
$$
\Eeins\big(\delta_\charge(F_1)+msF_1\big)\Omega= 
\Eeins (W_\charge+ms\unity)F_1\Omega=0
$$
by Eq.~\eqref{eqPauliLub1}. Lemma~\ref{BuEp} and the Reeh-Schlieder 
property~\eqref{eqReehSchlieEins} then imply that also 
\begin{equation} \label{eqF1Om0}
\EeinsConj\big((\delta_\charge(F_2))^\adj+msF_2^\adj\big)\Omega=0. 
\end{equation}
But by Eq.~\eqref{eqalphaConj}, the adjoint 
of $\delta_\charge(F_2)$ is $\delta_\concharge(F_2^\adj)$, and therefore 
$(\delta_\charge(F_2))^\adj\Omega = W_\concharge F_2^\adj\Omega$.
Then Eq.~\eqref{eqF1Om0} reads 
$$
\EeinsConj (W_\concharge+ms\unity)F_2^\adj\Omega=0. 
$$
This shows that only spin $s$ occurs in the single particle
space $\HeinsConj$, as claimed. 
The proof of the claim that not only the spin, but also the
multiplicity $n$ coincides then proceeds precisely as in \cite{BuEp}. 
\end{Proof}
\section{The Spin-Statistics Theorem} \label{SecSpiSta}
We now prove the spin-statistics theorem. Our line of reasoning
parallels that of Buchholz and Epstein~\cite{BuEp}, which uses heavily the 
representation of the covering of the Poincar\'e group. Since this
representation has completely different (analyticity) properties in
three dimensions, the corresponding details have to be
worked out differently in the present case.  

By our assumption (A3), the representation $U_\charge\big|{\Heins}$
is equivalent to $n$ copies of the irreducible representation of the
universal covering of the Poincar\'e group with mass $m>0$ and spin 
$s\in\RR$. Let us denote this representation by $U$. It acts on 
the Hilbert space $L^2(\Hyp,d\mu)\otimes \CC^n$,
elements of which  are functions (``wave functions'')
$$
\psi:\Hyp\times\{1,\ldots,n\}\to \CC, \quad
(p,\alp) \mapsto \psi(p,\alp) 
$$
with finite norm w.r.t.\ the scalar product 
$$
\big(\psi,\phi \big) =\int_{\Hyp}d\mu(p) \sum_{\alp=1}^n \; 
\overline{\psi(p,\alp)}\;\phi(p,\alp). 
$$
The representation $U$ acts in this space as 
\begin{equation} \label{eqUms} 
\big(U(a,\lortild)\psi\big) (p,\alp) = 
e^{is\Omega(\lortild,p)}\,e^{ia\cdot p}\,
\psi(\lor^{-1}p,\alp)\,, 
\end{equation}
where $\lor$ is the Lorentz
transformation onto which $\lortild$ projects, and 
$\Omega(\lortild,p)\in\RR$ is the Wigner rotation. The latter 
satisfies the so-called cocycle identities 
\begin{align} \label{eqCocyc}
\Omega(\unit,p) &=1, \qquad  
\Omega(\lortild\lortild',p) =
\Omega(\lortild,p)+\Omega(\lortild',\lor^{-1}p), 
\end{align}
and for the subgroup $\rottild{\cdot}$ of rotations (which is not
isomorphic to $SO(2)$ but to $\RR$) holds 
\begin{equation} \label{eqCocRot} 
\Omega\big(\rottild{\omega},p\big) = \omega \quad \text{ for all }
\;\omega\in\RR, \,p\in \Hyp.
\end{equation}
By Proposition~\ref{AntiPart}, $U_\concharge$ is also equivalent to
this representation. 
Thus, there are isometric isomorphisms $V_\charge$ and $V_\concharge$ 
from $\Heins$ and $\HeinsConj$ 
onto $L^2(\Hyp,d\mu)\otimes \CC^n$, which intertwine the 
representations $U_\charge\big|{\Heins}$ and 
$U_\charge\big|{\Heins}$, respectively, with $U$.

Following Buchholz and Epstein, we now fix two causally separated
(paths of) space-like cones 
$\spcpath_1$, $\spcpath_2$ as in the hypothesis of Lemma~\ref{BuEp},
and pick $n$ smooth field operators localized in either one of these
cones, $F_{i,\bet}\in\calF_\charge^\infty(\spcpath_i)$,
$\bet=1,\ldots,n$. We then consider, for $i=1$,
$2$, the wave functions 
\begin{align} \label{eqPsiibeta}
\psi_{i,\bet}&:= V_\charge\Eeins F_{i,\bet}\Omega  
& \text{ and } & & 
\psi^c_{i,\bet}&:= V_\concharge\EeinsConj F_{i,\bet}^\adj\Omega 
\intertext{in $L^2(\Hyp,d\mu)\otimes \CC^n$, and complex $n\times n$ 
 matrices $\Psi_i(p)$ and $\Psi^c_i(p)$ defined by }
\Psi_i(p)_{\alp\bet} &:= \psi_{i,\bet}(p,\alp) 
& \text{ and } && 
\Psi^c_i(p)_{\alp\bet} &:= \psi^c_{i,\bet}(p,\alp) 
\label{eqPsiMat}
\end{align}
for $p\in\Hyp$. 
We assume that the matrices $\Psi_i(p)$ are invertible for $p$ in
some open set on the mass shell. (This is possible due to the
Reeh-Schlieder property.) 
Lemma~\ref{BuEp} asserts that for each pair $\alp,\bet$ there is a
smooth function $h_{\alp\bet}$, analytic in $\Gamma$, such that 
\begin{align*}
h_{\alp\bet}(p)&=\sum_{\gam=1}^n\overline{\psi_{2,\alp}(p,\gam)}\,
\psi_{1,\bet}(p,\gam)
\equiv \big(\Psi_2(p)^\ast\,\Psi_1(p)\big)_{\alp\bet}\,,\\
h_{\alp\bet}(-p)&=\omega_\charge\;
\sum_{\gam=1}^n\overline{\psi^c_{1,\bet}(p,\gam)}\,
\psi^c_{2,\alp}(p,\gam)
\equiv 
\omega_\charge\;\big(\Psi^c_1(p)^\ast\,\Psi^c_2(p)\big)_{\bet\alp}, 
\end{align*}
where the star ${}^\ast$ denotes the matrix adjoint. 
(Note that this implies that the matrices $\Psi_i(p)$ and
$\Psi^c_i(p)$ are invertible for almost all $p$.) 
In other words, by Lemma~\ref{BuEp} the smooth matrix valued function
on the mass shell
\begin{equation} \label{eqMatAna}
p\mapsto \Psi_2(p)^\ast\,\Psi_1(p) =: M(p) 
\end{equation}
has an analytic extension into the subset $\Gamma$ of the
complexified mass shell described in 
\eqref{eqDomain}, with smooth boundary value on the negative mass
shell given by\footnote{The letter $p$ shall be reserved for points on
  the positive mass shell, so $-p$ is on the negative mass shell.}   
\begin{equation} \label{eqMatBV}
M(-p)= \omega_\charge\,\big(\Psi^c_1(p)^\ast\,\Psi^c_2(p)\big)^T, 
\end{equation}
where the superscript $T$ denotes matrix transposition. 
Buchholz and Epstein now proceed to show that, in the case of Bosons
and Fermions,  the wave function
matrices $\Psi_1(p)$ and $\Psi_2(p)^*$ separately have analytic extensions.  
This is not so in the present case. 
However, we show that their transforms under certain boosts behave 
analytically in the boost variable, which exhibits the underlying
modular covariance and is sufficient for our purpose. 

Let us recall the relevant geometric notions. 
We denote the one-parameter group of boosts in $x^1$-direction by
$\Boox{\cdot}$: 
\begin{equation}\label{eqBoost1}
\Boox{t} := 
\left( \begin{matrix} \cosh(t) & \sinh(t)&0 \\ \sinh(t) & \cosh(t)&0\\0&0&1 
\end{matrix}\right). 
\end{equation}
This matrix-valued function has an analytic extension into $\CC$ 
satisfying~\cite{H96} 
\begin{equation} \label{eqBooAna} 
 \Boox{t+i\theta} = \big(\J(\theta)+i \sin(\theta)\,
\sigma \big) \Boox{t}\,,
\end{equation} 
where $\J(\theta)=$ diag$(\cos\theta,\cos\theta,1)$  
and $\sigma$  maps $(x^0,x^1,x^2)$ to $(x^1,x^0,0)$. In particular, 
\begin{equation} \label{eqBooJ}  
\Boox{\pm i\pi}= \J,
\end{equation}
where $\J\equiv$ diag$(-1,-1,1)$ acts as the reflection of the coordinates 
$x^0$ and $x^{1}$, leaving $x^2$ unchanged. Note that $\J$ maps $\Hyp$
onto $H_m^-$ and satisfies $\J^2=\unity$. 

{}From now on we shall suppose that the dual of the 
``difference cone'' $\spcR_{12}$ contains the negative $x$-axis, that is: 
\begin{equation} \label{eqC12Neg}
\RR^-\times \{0\} \subset \big(\spcR_{12}\big)^*.
\end{equation}
In this case, for any $p\in\Hyp$ and any $z$ in the strip 
\begin{equation}\label{eqStrip} 
\strip:= \RR+i (0,\pi),
\end{equation}
the point $\Boox{-z}p$ is in the subset $\Gamma$ of the
complexified mass shell described in Lemma~\ref{BuEp}. (This is so because its
imaginary part is the image under $\sigma$ of a point in the past 
cone, hence of the form $(q_0,\boldsymbol{q})$ with
$\boldsymbol{q}\in\RR^-\times \{0\}$.) 
Hence by Lemma~\ref{BuEp} and Eq.~\eqref{eqMatAna}, for fixed $p
\in\Hyp$ the smooth matrix-valued function 
\begin{equation} \label{eqMatAnat'}
t\mapsto  \Psi_2(\Boox{-t}p)^\ast\,\Psi_1(\Boox{-t}p) \equiv  M(\Boox{-t}p) 
\end{equation}
has an analytic extension into the strip $\strip$, and, by 
Eqs.~\eqref{eqMatBV} and \eqref{eqBooJ}, its boundary value at $t=i\pi$ is  
\begin{equation} \label{eqMatAnaipit}
M(\Boox{-t}p)|_{t=i\pi} \equiv M(\J p)= 
\omega_\charge\,\big(\Psi^c_1(-\J p)^\ast\,\Psi^c_2(-\J p)\big)^T. 
\end{equation}
We need analyticity of $\Psi_1$ and $\Psi_2$ separately. However, it
turns out that it is not $\Psi_i(\Boox{-t}p)$ which is analytic, but
rather  the matrices $\Psi_i(t;p)$, $i=1,2$, defined  by 
\begin{align} \label{eqPsiMatt}
\Psi_i(t;p)_{\alp\bet}  :=&  \,
\big(U(\boox{t})\psi_{i,\bet}\big)(p,\alp) \\
 \equiv & \, e^{is\Omega(\boox{t},p)}\,\Psi_i(\Boox{-t}p)_{\alp\bet}. 
                  \label{eqPsiMatt'}
\end{align}
Here, $\boox{\cdot}$ denotes the unique lift to $\Lortild$ 
of the one-parameter group $\Boox{\cdot}$. 
The Wigner rotation factor in the last equation is independent  
of $\alp,\bet$ and $i$,
and therefore cancels in Eq.~\eqref{eqMatAnat'}. Hence
Eq.~\eqref{eqMatAnat'} implies that 
\begin{equation} \label{eqMatAnat}
t\mapsto  \Psi_2(t;p)^\ast\,\Psi_1(t;p) \equiv  M(\Boox{-t}p) 
\end{equation}
has an analytic extension into the strip $\strip$ with boundary value 
given by Eq.~\eqref{eqMatAnaipit}. 
\begin{Lem} \label{AnaFun}
For any $p\in\Hyp$, the smooth matrix-valued functions 
$t\mapsto \Psi_1(t;p)$ and $t\mapsto
\Psi_2(t;p)^*$ extend to analytic functions on the strip $\strip$ with
smooth boundary values at the upper boundary $\RR+i\pi$. 
\end{Lem}
(Note that $\Psi_2(t;p)$ is analytically continued {\em after} conjugation.)   
\begin{Proof}
The proof uses the same reasoning as~\cite[Sect.~3]{BuEp}. 
Let us denote, for brevity, $f_1(t):=\Psi_1(t;p)$,
$f_2(t):=\Psi_2(t;p)^\ast$ and $h(t):=M(\Boox{-t}p)$. We know,
by Eq.~\eqref{eqMatAnat}, that $t\mapsto f_2(t)f_1(t)\equiv h(t)$ has
an analytic extension into the strip $\strip$. 
The equation 
$$
f_1(t+t_0)_{\alp\bet}=
\big(U(\boox{t}) V_\charge\Eeins
\alpha_\charge(\boox{t_0})(F_{1,\bet})\Omega\big) (p,\alp)
$$
shows that$f_1(t+t_0)$ is of the same form as $f_1(t)$, 
with $F_{1,\bet}$ substituted by 
$\alpha_\charge(\boox{t_0})(F_{1,\bet})$. 
Now for $t_0$ sufficiently small, $\boox{t_0}\act\spcpath_1$ still
satisfies (together with $\spcpath_2$) the hypothesis of
Lemma~\ref{BuEp} and condition~\eqref{eqC12Neg}. Hence, the same
reasoning as above shows that there is a matrix-valued function 
$h_{t_0}(t)$ analytically extendible in $t$ into the strip $\strip$, such that 
\begin{equation} \label{eqMatAnat''}
f_2(t)f_1(t+t_0)= h_{t_0}(t)
\end{equation}
for $t_0$ sufficiently small. Smoothness of $F_{1,\bet}$ implies that 
$f_1$ is smooth and that $h_{t_0}(t)$ is smooth in $t_0$. The above
equation implies that 
\begin{equation} \label{eqff'}
f_2(t)\, \frac{{\rm d}}{{\rm d}t} f_1(t) 
= \hat h(t):= \frac{{\rm d}}{{\rm d}t_0} h_{t_0}(t)
\big|_{t_0=0} .
\end{equation}
The last two equations imply the following differential equation for $f_1$: 
\begin{equation} \label{eqf-1f'}
f_1(t)^{-1}\, \frac{{\rm d}}{{\rm d}t} f_1(t)=h(t)^{-1}\,\hat h(t). 
\end{equation}
The right hand side is meromorfic in the strip $\strip$ and continuous
on its closure $\strip^\clo$ (up to isolated points). Hence $f_1$
can be integrated along any path $\gamma$ in $\strip^\clo$ starting
from the real (=lower) boundary, as long as the path does not cross 
zeroes of the
determinant of $h(z)$, yielding an analytic extension $f_{1,\gamma}$
along $\gamma$. If $\gamma$ crosses a zero $z_0$ of det $h(z)$, we
make use of the following observation: Eq.~\eqref{eqMatAnat''} implies
the relation 
\begin{equation} \label{eqf1hh}
f_1(t)=f_1(t+t_0)\,h(t+t_0)^{-1}\, h_{-t_0}(t+t_0), 
\end{equation}
which extends from real $t$ to values in the strip $\strip$, along the 
path $\gamma$. 
Since the zeroes of det $h(z)$ are isolated, the determinant of 
$h(z_0+t_0)$ is non-zero for $t_0$ sufficiently small. 
Thus, the function $f_{1,\gamma}$ can be 
continuously (and hence analytically) continued into $z_0$ by the
(analytic extension of the) above equation. Hence, $f_1$ extends
analytically along any path into the strip. But the latter is simply connected,
hence the analyic extensions are independent of the paths, proving the
claimed analyticity of $t\mapsto \Psi_1(t;p)$. Smoothness of the
boundary value at $\RR+i\pi$ follows from Eq.~\eqref{eqf1hh}. 
Analyticity of $\Psi_2(t;p)^*\equiv f_2(t)$ is shown along the same lines. 
\end{Proof}
Lemma~\ref{AnaFun} allows for the definition of ``geometric
Tomita operators'' acting on the matrix-valued functions $\Psi_{1}$ and
$\Psi_{2}$. Namely, we define for $p\in \Hyp$ 
\begin{align}\label{eqTomop}
\hat\Psi_1(p) := \overline{\Psi_1(t;-\J p)|_{t=i\pi}},\quad 
\check\Psi_2(p) := \Big(\overline{\Psi_2(t;-\J p)}\Big)\big|_{t=i\pi}, 
\end{align}
where complex conjugation is understood componentwise. 
(Note that $\Psi_1$ is first analytically continued to $t=i\pi$ and
then conjugated, while $\Psi_2$ is first conjugated and then
continued.) 
We now have 
\begin{align*} 
 \hat\Psi_1(p)^\ast\,\check\Psi_2(p) = 
\big\{\Psi_2(t;-\J p)^*\Psi_1(t;-\J p)\big\}^T\big|_{t=i\pi}
\end{align*}
by definition. But the function in curly brackets coincides, by
Eq.\ \eqref{eqMatAnat}, with $M(-\Boox{-t}\J p)$ whose analytic
continuation into $t=i\pi$ is $M(-p)$ by Eq.~\eqref{eqBooJ}. Using 
Eq.~\eqref{eqMatAnaipit}, we therefore have 
\begin{equation} \label{eqMatAnaTom}
 \hat\Psi_1(p)^\ast\,\check\Psi_2(p) = 
\omega_\charge\,\Psi^c_1(p)^\ast\,\Psi^c_2(p).
\end{equation}
We want to find a relation between $\hat\Psi_1$ and $\Psi^c_1$,
constituting a Bisognano-Wich\-mann property on the single particle
level~(Proposition~\ref{Tom}). The proof of this relation relies on
the fact that the matrix-valued function 
$\hat\Psi_1$ transforms under Lorentz transformations (close to unity)
just like $\Psi_1$ (Lemma~\ref{Transform}). The proof of this
transformation behaviour is the crucial and
difficult point in our analysis, since the Wigner rotation factor
spoils  the analyticity needed for the definition of $\hat\Psi_1$. 
Observe that for $\lor\in\Lor$ sufficiently small,
$\lor\spc_1$ is contained in a space-like cone of the form 
 $(\spcR_1^\lor)''$, 
which satisfies, together
with $\spc_2$, the hypothesis of Lemma~\ref{BuEp} and the
condition~\eqref{eqC12Neg},   
$\RR^-\times\{0\}\subset (\spcR_2-\spcR_1^\lor)^*$. 
Let $\calU_{12}$ be a neighbourhood of the identity in $\Lor$ consisting of
such $\lor$. The set of $\lortild\in\Lortild$ which project onto
$\calU_{12}$ has an infinity of connected components,
differing by $2\pi$-rotations. Let now $\calUtild_{12}$ be the one 
containing the identity. This ensures that for
$\lortild\in\calUtild_{12}$, the paths $\lortild\act\spcpath_1$
and $\spcpath_2$ have the correct relative winding
number so as to satisfy the hypothesis of Eq.~\eqref{eqCommut}.   
Then, for $\lortild \in \calUtild_{12}$, the wave function\footnote{We
  use a superscript $\lor$ instead of $\lortild$, which causes no 
  confusion since we have a one-to-one correspondence between $\calU_{12}$ and
  $\calUtild_{12}$.}  
\begin{equation} \label{eqPsiLor}
\psi_{1,\beta}^{\lor}:=U(\lortild)\psi_{1,\bet} \equiv V_\charge \Eeins
\alpha_\charge(\lortild)(F_{1,\bet}) \,\Omega
\end{equation}
is of the same form as $\psi_{1,\bet}$, with $F_{1,\bet}$ substituted by 
$\alpha_\charge(\lortild)(F_{1,\bet})$, and 
Lemma~\ref{AnaFun} applies, asserting that the matrix-valued function 
$$
t\mapsto \Psi_1^\lor(t;p)_{\alp\bet}:=
\big(U(\boox{t})\psi_{1,\bet}^\lor\big)(p,\alp) 
$$
has an analytic extension into $\strip$, with continuous boundary value
at $\RR+i\pi$. This  allows for the definition of 
\begin{align}\label{eqTomopLor}
\widehat{\Psi_1^{\lor}}(p) := \overline{\Psi_1^\lor(t;-\J p)|_{t=i\pi}}, 
\end{align}
in analogy with Eq.~\eqref{eqTomop}.  
\begin{Lem} \label{Transform} There is a neighbourhood $\calUtild$ of
the unit in $\Lortild$ such that for all $\lortild\in \calUtild$ and
$p\in\Hyp$ there holds  
\begin{align} \label{eqTransformLor}
\widehat{\Psi_1^{\lor}}(p)&= e^{is\Omega(\lortild,p)}\,
\hat\Psi_1(\lor^{-1}p)\,.
\end{align}
\end{Lem}
\begin{Proof}
The claimed equation is equivalent with 
\begin{multline} \label{eqTransformLor'}
e^{is\Omega(\boox{t}\lortild,-\J p)}\,
\psi_{1,\bet}(-\lor^{-1}\Boox{-t}\J p,\alp)\big|_{t=i\pi} = \\ 
e^{-is\Omega(\lortild,p)}\, e^{is\Omega(\boox{t},-\J \lor^{-1}p)}\,
\psi_{1,\bet}(-\Boox{-t}\J \lor^{-1}p,\alp)\big|_{t=i\pi}.
\end{multline}
Now the function $t\mapsto e^{is\Omega(\boox{t}\lortild,q)}$ has
branch points in the strip $\strip$, see Lemma C.1 of \cite{M02a}. Hence
none of the ($t$-dependent) factors in the above equation
possesses an analytic extension into the strip by its own.  
However, we have constructed in~\cite{M02a} a function living
on the mass shell which compensates the singularities of the Wigner 
rotation factor. In Appendix~\ref{Cocycles}, we adopt the results
of~\cite{M02a} to the present situation, leading to the following
assertion (c.f.\ Lemma~\ref{CocW1'}). Let 
\begin{align} \label{equPi2'}
u_{\pihalf}(p) &:=
e^{is\pihalf} \,  \Big(\frac{p_0-p_2}{m}\cdot 
\frac{{p_0-p_2+m+ip_1}}{p_0-p_2+m-ip_1}\Big)^{s}\,\quad \text{ and }\\
\omega(\lortild,p) &:= e^{is\Omega(\lortild,p)}\,u_{\pihalf}(\lor^{-1}p).
\end{align}
Then the function $t\mapsto \omega(\boox{t}\lortild,p)$ 
has an analytic extension into the strip $\strip$ for all
$\lortild$ in a neighbourhood $\calUtild_0$ of the unit. Further, at
$t=i\pi$ it has the boundary value 
\begin{align} \label{eqBV} 
\omega(\boox{i\pi}\lortild,p)&=
e^{i\pi s}\, e^{is\Omega(\J \lortild \lortild_0\J ,p)}\, 
u\big(\J  (\lor \lor_0)^{-1}\J p\big),\;  \text{ where } \\ 
u(p)&:= \Big(\frac{p_0-p_1}{m}\cdot
\frac{{p_0-p_1+m-ip_2}}{p_0-p_1+m+ip_2}\Big)^{s}. 
\end{align}
Here, $\lor_0:=\rot{\pi/2}$ is the rotation about $\pi/2$, and 
$\lortild_0:=\rottild{\pi/2}$ where $\rottild{\cdot}$ is the unique
lift to $\Lortild$ of the one-parameter group of rotations. Further, 
$\lortild\mapsto \J \lortild \J $ is the unique lift~\cite{Var2} of
the adjoint action of $\J $ on $\Lor$ to an automorphism of the
universal covering group. 
To apply this result, we rewrite the claimed 
equation~\eqref{eqTransformLor'} as follows:  
\begin{multline}    \label{eqTransformLor''} 
\omega(\boox{t}\lortild,-\J p) \cdot  
\phi(-\lor^{-1}\Boox{-t}\J p)\,\big|_{t=i\pi}  \\
= \big(e^{-is\Omega(\lortild,p)}\,
\omega(\boox{t},-\J \lor^{-1}p)\big)  \cdot 
\phi(-\Boox{-t}\J \lor^{-1}p)\,\big|_{t=i\pi} ,
\end{multline}
where 
\begin{align} \label{eqAnaCoc}
\phi(p):= u_{\pihalf}(p)^{-1}\, \psi_{1,\bet}(p,\alp). 
\end{align}
Lemma~\ref{CocW1'} then asserts that for $\lortild\in \calUtild_0$ the
first factor $\omega(\boox{t}\lortild,-\J p)$ on the left hand
side of Eq.~\eqref{eqTransformLor''} is
analytic in $\strip$ and has the boundary value 
\begin{equation} \label{eqBVLHS} 
e^{i\pi s}\, e^{-is\Omega(\lortild \lortild_0,p)}\, 
u\big(-\J  (\lor \lor_0)^{-1}p\big)  
\end{equation}
at $t=i\pi$. (Here we have used that the Wigner rotation satisfies the
identity
\begin{align}  \label{eqCocycjj} 
\Omega(\J \lortild \J ,p) & = -\Omega(\lortild,-\J p), 
\end{align}
see~\cite[Lemma B.2]{M02a}.) Similarly, the first factor  
$e^{-is\Omega(\lortild,p)}\, \omega(\boox{t},-\J \lor^{-1}p)$ 
on the right hand side of Eq.~\eqref{eqTransformLor''} is analytic,
with boundary value 
\begin{equation} \label{eqBVRHS} 
e^{i\pi s}\, e^{-is\Omega(\lortild,p)}\, 
e^{-is\Omega(\lortild_0,\lor^{-1}p)}\, u\big(-\J  (\lor \lor_0)^{-1}p\big)  
\end{equation}
at $t=i\pi$. Due to the cocycle identity~\eqref{eqCocyc}, 
this coincides with the boundary value~\eqref{eqBVLHS} of the first
factor on the left hand side of Eq.~\eqref{eqTransformLor''}. 

We now know that for any $\lortild\in\calUtild:=\calUtild_0\cap
\calUtild_{12}$ both sides of Eq.~\eqref{eqTransformLor''}
are analytic in the strip $\strip$, and the same holds for the first
factor on each side. Further, we know that the boundary values 
at $t=i\pi$ of the first factors coincide. 
It follows that the second factors, namely the functions 
\begin{equation} \label{eqPhi}
f_1(t)= \phi(-\lor^{-1}\Boox{-t}\J p)\qquad \text{ and } \qquad 
f_2(t)= \phi(-\Boox{-t}\J \lor^{-1}p), 
\end{equation}
also have an analytic extension into the strip. It only remains
to show that their boundary values at $t=i\pi$ coincide. 
To this end, note that the analyticity of the two
functions~\eqref{eqPhi} holds for all $p\in \Hyp$ and
$\lor$ in the projection of $\calUtild$ onto $\Lor$, which we shall
denote by $\calU$. Hence we can analytically continue the function
$\phi$ into the subset 
$$
\Gamma_0:=\{ \lor \Boox{z} p: \; p\in\Hyp,z\in\strip, \lor\in\calU\}
$$
of the complexified mass shell $\HypC$ along paths of the form
$\lor\Boox{z(t)}p$. Now a straightforward calculation shows that every
$k=\lor \Boox{z} p\in\Gamma_0$ can be uniquely written in the form 
$k=r\Boox{i\theta}r^{-1}q$, where $r$ is a rotation,
$\theta\in(0,\pi)$ and $q\in\Hyp$. By restricting
$\lor$ to a smaller neighbourhood if necessary, one can achieve 
$r\in\calU$. Letting $\theta$ go to zero then defines a deformation
retraction of $\Gamma_0$ onto the mass hyperboloid. Hence
$\Gamma_0$ is simply connected, which implies that our analytic
continuation of $\phi$ is path-independent, yielding an analytic
function $\hat \phi$ on $\Gamma_0$, continous at the real boundary $H_m^{-}$,  
such that $f_1(z)=\hat \phi(-\lor^{-1}\Boox{-z}\J p)$ and 
$f_2(z)= \hat \phi(-\Boox{-z}\J \lor^{-1}p)$.  
But the points $-\lor^{-1}\Boox{-i\pi}\J p$ and 
$-\Boox{-i\pi}\J \lor^{-1}p$ coincide, namely with $-\lor^{-1}p$,
hence $f_1(i\pi)=f_2(i\pi)$. This completes the proof.
\end{Proof}

\begin{Prop} \label{Tom}
The following ``Bisognano-Wichmann property'' holds: 
There is a regular $n\times n$ matrix $D$ such that for all $p\in\Hyp$ there
holds 
\begin{equation}\label{eqTom} 
\hat\Psi_1(p) = D\, \Psi^c_1(p). 
\end{equation}
\end{Prop}
\begin{Proof} 
The proof goes again along the lines of~\cite{BuEp}, but uses our
Lemma~\ref{Transform}.  
Let $p$ be in the dense set of points satisfying $\det \Psi^c_1(p)\neq0$,
and let $D(p)$ be the matrix
$$
D(p):= \hat\Psi_1(p)\, \Psi^c_1(p)^{-1}. 
$$
Due to Eq.~\eqref{eqMatAnaTom}, $D(p)$ is independent of the specific
choice of operators $F_{1,\bet}$ from which $\hat\Psi_{1}(p)$ and
$\Psi^c_1(p)$ are constructed. In particular, for 
$\lortild\in \calUtild_{12}$, we may substitute 
$F_{1,\bet}$ by $\alpha_\charge(\lortild)(F_{1,\bet})$ as
in Eq.~\eqref{eqPsiLor}, yielding substitution of $\hat\Psi_{1}(p)$ by
$\widehat{\Psi_1^{\lor}}(p)$ and of  $\Psi^c_1(p)_{\alp\bet}$ by 
$$
\Psi_{1}^{\lor,c}(p)_{\alp\bet} := 
\big( U(\lortild)\, V_\concharge \EeinsConj F_{1,\bet}^\adj
\,\Omega\big)(p,\alp).
$$
Hence we have 
\begin{align*} 
D(p)&= \widehat{\Psi_1^{\lor}}(p)\, \Psi^{\lor,c}_1(p)^{-1} =
\hat\Psi_1(\lor^{-1}p)\, \Psi^c_1(\lor^{-1}p)^{-1} = D(\lor^{-1}p).
\end{align*}
(In the second equation we have used that, by Lemma~\ref{Transform},   
$\widehat{\Psi_1^{\lor}}(p)$ and $\Psi^{\lor,c}_1(p)$ have the same
transformation dependence on $\lortild$, namely 
$\Psi_1^{\lor,c}(p)=e^{is\Omega(\lortild,p)}$$\Psi_1^c(\lor^{-1}p)$
and Eq.~\eqref{eqTransformLor}.) This shows that $D(p)$ is locally
constant, and, since $p$ was arbitrary, constant.  
\end{Proof}
As a corollary, we get a relation between $\check\Psi_2(p)$ and 
$\Psi^c_2(p)$.  
\begin{Cor} \label{Tom'}
For all $p\in\Hyp$ there holds 
\begin{equation}\label{eqTom'} 
\check\Psi_2(p) = e^{2\pi i s}\, D\, \Psi^c_2(p). 
\end{equation}
\end{Cor}
\begin{Proof} 
Let us choose our paths $\spcpath_1$ and $\spcpath_2$ so as to satisfy 
$\spcpath_1=\rottild{\pi}\act \spcpath_2$, where $\rottild{\cdot}$ 
denotes the one-parameter group of rotations in $\Lortild$. 
(This is compatible with the hypothesis of Lemma~\ref{BuEp}.) 
Then the wave function 
\begin{align} \label{eqPsiPi} 
\psi_{2,\bet}^{\pi} & := U(\rottild{\pi})\psi_{2,\bet}\equiv 
V_\charge \Eeins \alpha_\charge(\rottild{\pi})(F_{2,\bet})\,\Omega
\end{align}
is of the same form as $\psi_{1,\bet}$, with
$F_{1,\bet}$ substituted by
$\alpha_\charge(\rottild{\pi})(F_{2,\bet})$. Hence,
Lemma~\ref{AnaFun} allows for the analytic extension 
$$
\widehat{\Psi_2^{\pi}}(p)_{\alp\bet} :=
\overline{\big(U(\boox{t})\psi_{2,\bet}^{\pi}\big)(-\J p,\alp)|_{t=i\pi}}.
$$
Now the group relation $\boox{t}\rottild{\pi}=\rottild{\pi}\boox{-t}$ 
implies that 
\begin{multline}
\widehat{\Psi_2^{\pi}}(p)_{\alp\bet} = 
\overline{\big(U(\rottild{\pi})U(\boox{-t})
\psi_{2,\bet}\big)(-\J p,\alp)|_{t=i\pi}} \\
\equiv e^{-i\pi s}\, \overline{\big(U(\boox{-t})
\psi_{2,\bet}\big)(-\rot{-\pi}\J p,\alp)|_{t=i\pi}}.  
\end{multline}
(In the last equation we have used relation~\eqref{eqCocRot}.) 
The group relation $\rot{-\pi}\J = \J \rot{\pi}$ and the
identity $\overline{f(-t)|_{t=i\pi}}=\bar f(t)|_{t=i\pi}$,
holding  for the analytic extension of a function $\bar f$, yield 
\begin{align} \label{eqTransformPi}
\widehat{\Psi_2^{\pi}}(p)&= e^{-is\pi}\, \check\Psi_2(r(\pi)p)\,.
\end{align}
On the other hand, Proposition~\ref{Tom} asserts that 
\begin{align} \label{eqTransformPi'}  
\widehat{\Psi_2^{\pi}}(p)= 
D\, \Psi^{\pi,c}_2(p),  
\end{align} 
where $\Psi^{\pi,c}_2(p)$ is defined just as $\Psi^{c}_1(p)$ with
$F_{1,\beta}^\adj$ substituted by 
$\alpha_\concharge(\rottild{\pi})(F_{2,\bet}^\adj)$.  
But using Eq.~\eqref{eqCocRot} yields $\Psi^{\pi,c}_2(p)=\exp(i\pi s)\,
\Psi^{c}_2(\rot{-\pi}p)$.  Hence, taking into account that
$\rot{\pi}=\rot{-\pi}$, Eqs.~\eqref{eqTransformPi} and
\eqref{eqTransformPi'} imply the claimed Equation~\eqref{eqTom'}. 
\end{Proof}
This implies our main result, the relation between spin and  
statistics for anyons and plektons:
\begin{Thm}[Spin-Statistics Theorem] \label{SpiSta}
The spin $s$ and statistics phase $\omega_\charge$ are related by  
$$
e^{2 \pi i s}=\omega_\charge. 
$$
\end{Thm}
\begin{Proof} 
Substituting Eqs.~\eqref{eqTom} and \eqref{eqTom'} into 
Eq.~\eqref{eqMatAnaTom}, yields 
$$
D^\ast D\, e^{2\pi i s} = \omega_\charge \unity,  
$$ 
since the matrices $\Psi^c_i(p)$ are invertible for almost all $p$. 
Uniqueness of the polar decomposition then implies the claim.  
\end{Proof}
\appendix 
\renewcommand{\theequation}{\thesection.\arabic{equation}}
\renewcommand{\theLem}{\thesection.\arabic{Lem}}
\section{Justification of the Assumptions} 
\label{Just}
\setcounter{equation}{0}
\setcounter{Lem}{0}
We assume the standard assumptions on the algebra $\calA$ of local
observables~\cite{H96} plus weak Haag Duality of the vacuum 
representation~\cite[Eq.~(1.11)]{BuF}, 
and consider a covariant representation
$\pi_\charge$ of $\calA$ which is strictly massive in the sense of our 
assumptions (A1) and (A2). As shown in~\cite{BuF}, $\pi_\charge$ is
then localizable in space-like cones, i.e., equivalent to the  vacuum
representation when restricted to the causal complement of a space-like
cone. One can then enlarge the algebra of observables to the so-called
universal algebra $\Auni$~\cite{FRSII,FGR} and find an endomorphism
$\varrho$ of $\Auni$ such that the (unique lift of the) representation 
$\pi_\charge$ is equivalent to the representation $\pi_0\circ\varrho$,
where $\pi_0$ is the vacuum representation of $\Auni$ acting in a
vacuum Hilbert space $\HVac$. 
The endomorphism $\varrho$ is localized in some specific space-like
cone $\spc_0$ in the sense that 
\begin{equation} \label{eqRhoInC0}
\varrho(A)=A\qquad \text{ if } \; A \in\Auni(C_0'),
\end{equation}
where $C_0'$ denotes the causal complement of $C_0$. 
The endomorphism $\varrho$ has a conjugate $\bar \varrho$ such
that $\bar{\varrho}\varrho$ contains the identity representation
$\iota$ of $\Auni$~\cite{BuF}. We shall choose a corresponding
intertwiner $R\in\Auni$. 
\footnote{In \cite{DHRIV} a different normalization convention is used,
  namely $R^*R=|\lambda_\chi|^{-1}\unity$~\cite[Eq.~(3.14)]{DHRIV}.}    
Associated with $\varrho$ is the statistics operator $\eps_\varrho$,
which describes the interchange of two charges localized in causally
separated space-like cones. Using the notions of our
Section~\ref{SecAss}, it is constructed as follows. 
We fix the reference direction $\spd_0$ so as to be contained, in the sense of
Eq.~\eqref{eqDirInSpc}, in $\spc_0$. Let $\spcpath_1=(\spc_1,\spdpath_1)$ and
$\spcpath_2=(\spc_2,\spdpath_2)$ be paths of space-like cones
satisfying the hypothesis of Eq.~\eqref{eqCommut}. Let further 
$U_i$, $i=1,2$, be (heuristically speaking) charge transporters which 
transport the charge $\varrho$ from  $\spc_0$ to $\spc_i$ along the 
path $\spdpath_i$. This means the following. $U_i$ is an intertwiner such 
that $\text{Ad }U_i\circ \varrho$ is localized in $\spc_i$ (instead of
$\spc_0$) in the sense of
Eq.~\eqref{eqRhoInC0}, and at the same time is an observable localized in
$I_i$, where $I_i$ a space-like cone (or the complement of one)
containing the complete path $\spdpath_i(t)$, $t\in[0,1]$, in the sense of
Eq.~\eqref{eqDirInSpc}. Then 
\begin{equation} \label{eqEps}
\eps_\varrho:= \varrho(U_1^*)U_2^*U_1\varrho(U_2).
\end{equation}
The corresponding statistics parameter $\lambda_\charge$ and statistics phase
$\omega_\charge$ are then defined by the relations  
\begin{equation} 
\phi(\eps_\varrho)=\lambda_\charge \unity, \quad \omega_\charge =
\frac{\lambda_\charge}{|\lambda_\charge|},  
\end{equation}
respectively. (They depend only on the equivalence class of $\varrho$,
i.e., on its sector $\charge$.)  
Here, $\phi$ is the left inverse of $\varrho$, that is 
a positive linear endomorphism of $\Auni$ satisfying 
\begin{equation} \label{eqLeftInv}
\phi\big(\varrho(A)B\varrho(C)\big)=A\phi(B)C,\quad \phi(\unity)=\unity.
\end{equation}
It can be expressed as~\cite{DHRIV,BuF} 
\begin{equation} \label{eqLeftInvR}
\phi(A)= 
R^*\bar\varrho(A) R.
\end{equation}
We now identify the objects and notions of our Section~\ref{SecAss}
within the frame indicated above and with objects derived within this
framework in~\cite{DHRIV,BuF,FRSII,FGR}.  
Our sectors $\charge$ and $\concharge$ are just the  equivalence classes of the
representations $\pi_0\circ\varrho$  and $\pi_0\circ\bar\varrho$,
respectively. Our Hilbert spaces $\calH_0$, $\calH_\charge$ and
$\calH_\concharge$ are the fibres $\{\iota\}\times \HVac$, 
$\{\varrho\}\times \HVac$ and $\{\bar\varrho\}\times \HVac$ of the
vector bundle $\calH$ of generalized state vecors introduced in~\cite{DHRIV},
see also~\cite{BuF}, respectively. 
The respective scalar products are inherited by that of $\HVac$. 
Our vacuum vector $\Omega$ is identified with the Poincar\'e invariant
vector $\Omega_0$ inducing the vacuum state: 
$$
\Omega=(\iota,\Omega_0) \; \in \,\calH_0. 
$$    
The spaces of our fields $\calF_\charge$ and $\calF_\concharge$ are
defined as the subspaces $\{\varrho\}\times  \Auni$ and 
$\{\bar\varrho\}\times\Auni$, respectively, of the field bundle  $\calF$ 
introduced in~\cite{DHRIV}.  
A generalized field operator $F=(\varrho,B)\in\calF_\charge$ then acts on a
generalized state vector $(\iota,\psi)\in\calH_0$ as 
$$
(\varrho,B)\, (\iota,\psi):=(\varrho,\pi_0(B)\psi) \;\in\calH_\charge. 
$$
The adjoint $F^\adj$ of a generalized field operator
$F=(\varrho,B)\in\calF_\charge$ is defined by 
\begin{equation} \label{eqFadj}
(\varrho,B)^\adj:= |\lambda_\charge|^{-\half}\,
 (\bar\varrho,\bar\varrho(B^*)R),
\end{equation}
where $B^*$ is the $C^*$-adjoint of $B$ in $\Auni$. 

The notion of localized generalized field operators has been
introduced in~\cite{DHRIV} in the case of permutation group
statistics. The extension to the case of braid group statistics needs a 
refinement, which has been introduced in~\cite{FRSII}, see also
\cite{FGR}. There, $\calK$ denotes the class of space-like cones or
causal complements  thereof, and a path in $\calK$ is a finite sequence
$(I_0,\ldots,I_n)$, $I_k\in\calK$, such that either $I_k\subset I_{k-1}$ or 
$I_k\supset I_{k-1}$, $k=1,\ldots,n$. We say that such path starts at $C_0$ if
$I_0=C_0$. 
The relation to our notion of paths of space-like cones,
Eq.~\eqref{eqSpcpath}  is as follows.   
Our $(\spc,\spdpath)$ corresponds to a path 
$(I_0,\ldots,I_n)$ in $\calK$ starting at $\spc_0$ if $\spdpath$,
considered as a path in $\Spd$, has the decomposition
$\spdpath=\gamma_n\ast\cdots\ast\gamma_0$ such that $\gamma_k(t)$ is
contained in $I_k$ in the sense of Eq.~\eqref{eqDirInSpc} for all
$t\in[0,1]$ and $k=0,\ldots,n$.  With this identification, our space 
of localized fields $\calF_\charge(\spcpath)$ is defined as 
$$
\calF_{\charge}(\spcpath):= \calF_\charge \cap \calF(\spcpath), 
$$
where $\calF(\spcpath)$ is the space of generalized field operators
localized along $\spcpath$ as defined in \cite{FRSII,FGR}.
$\calF_\concharge(\spcpath)$ is defined analogously.  
The fact that the adjoint preserves localization,
Eq.~\eqref{eqFadjLoc}, is just Eq.~(6.37) in~\cite{BuF} (which
strengthens Lemma~4.3 in~\cite{DHRIV}).  

Our representations $U_\charge$ and $\alpha_\charge$ of the universal 
covering group of the Poincar\'e group in $\calH_\charge$ and
$\calF_\charge$, respectively, are defined as follows. 
Let $U(\potild)$ and 
$\alpha(\potild)$ be the representations in $\calH$ and
$\calF$ as defined in 
\cite[Eqs.~(4.3) and (4.4)]{DHRIV} in the case of permutation group 
statistics, and \cite[Eqs.~(2.18) and (2.19)]{FGR} in the case of
braid  group statistics, respectively. 
Then we define 
$$
U_\charge(\potild):= U(\potild)\big|\calH_\charge\quad \text{ and } 
\quad 
\alpha_\charge(\potild):=\alpha(\potild)\big|\calF_\charge. 
$$
The covariance condition~\eqref{eqCov} is just Eq.~(4.7)
in~\cite{DHRIV}. 
Our Eq.~\eqref{eqalphaConj}, relating the adjoint,
$\alpha_\charge$ and $\alpha_{\concharge}$  (defined analogously), is
just Eq.~(4.20) in \cite{DHRIV}. 
The fact that Eqs.~\eqref{eqCov},  \eqref{eqFadjLoc} and
\eqref{eqalphaConj} also hold in the case of braid group statistics
has been shown in \cite{MDiss}.  

Our Eq.~\eqref{eqCommut}, fixing the significance of the statistics
phase $\omega_\charge$, corresponds to Eq.~(6.5) in \cite{DHRIV} in
the case of permutation group statistics. But 
since we are not aware of literally the same equation in the
literature in the case of braid group statistics, we give a direct
proof, transferring their arguments to this case. 
\begin{Lem} 
Let $\spcpath_1=(\spc_1,\spdpath_1)$ and
$\spcpath_2=(\spc_2,\spdpath_2)$ be paths of space-like cones 
satisfying the hypothesis of Eq.~\eqref{eqCommut}. Let further
$F_i=(\varrho,B_i)\in\calF_\charge(\spcpath_i)$, $i=1,2$.   
Then there holds Eq.~\eqref{eqCommut}, namely, 
$$
\big(\, F_2\Omega,F_1\Omega\,\big) =  
\omega_\charge \,\big(\, F_1^\adj\Omega,F_2^\adj\Omega\,\big).
$$
\end{Lem}
\begin{Proof}
$(\varrho,B_i)\in\calF_\charge(\spcpath_i)$ means that there are 
unitary charge transporters $U_i$ satisfying presisely the
hypothesis of Eq.~\eqref{eqEps}, and that 
$A_i:=U_iB_i$ is an observable localized in $\spc_i$, $i=1,2$. 
Denoting $\varrho_i:=\text{Ad }U_i\circ \varrho$, we then have 
\begin{align} 
\varrho(B_2^*)\,\eps_\varrho^*\,\varrho(B_1) &=  
\varrho(B_2^*U_2^*)\,U_1^*\,U_2\varrho(U_1B_1)= 
\varrho(A_2^*)\,U_1^*\,U_2\varrho(A_1)\label{i} \\
&= U_1^*\varrho_1(A_2^*)
\varrho_2(A_1)U_2 = U_1^* A_2^*A_1U_2 = U_1^*A_1A_2^*U_2=B_1\,B_2^*. \nonumber
\end{align}
(We have used that $\varrho_i$ are localized in $\spc_i$ in the sense of
Eq.~\eqref{eqRhoInC0} and that $A_1$ and $A_2^*$ commute due to
locality of the observables.)  
Applying the left inverse $\phi$ to 
Eq.~\eqref{i}, using the explicit formula~\eqref{eqLeftInvR} for the 
left inverse and taking into account that $\phi$ preserves
the $C^*$-adjoint, yields 
\begin{align*}
\bar\lambda_\charge \,B_2^*\,B_1 &= \; R^*\bar\varrho(B_1B_2^*)R\,. 
\end{align*}
Using this equation, we get 
\begin{align*} 
\big(\, F_2\Omega,F_1\Omega\,\big) & = 
\big(\, \Omega_0,\pi_0(B_2^*\,B_1)\,\Omega_0\,\big) =
(\bar\lambda_\charge)^{-1}\;  
\big(\, \Omega_0, \pi_0(R^*\bar\varrho(B_1B_2^*)R)  \Omega_0\,\big) \\
& =
(\bar\lambda_\charge)^{-1} \;  
\big(\, \pi_0(\bar\varrho(B_1^*)R) \Omega_0,
\pi_0(\bar\varrho(B_2^*)R)  \Omega_0\,\big) =
\omega_\charge \,\big(\, F_1^\adj\Omega,F_2^\adj\Omega\,\big),  
\end{align*}
since $(\bar\lambda_\charge)^{-1}|\lambda_\charge| = \omega_\charge$.  
This completes the proof. 
\end{Proof}
\section{An Analytic Cocycle for the Massive Irreducible
  Representations of $\boldsymbol{\Potild}$ 
in 2+1 Dimensions} \label{Cocycles} 
\setcounter{equation}{0}
\setcounter{Lem}{0}
In~\cite{M02a}, we have shown that the Wigner rotation factor 
$\exp (is\Omega(\lortild,p))$ is
non-analytic in the sense that the function 
$t\mapsto \exp (is\Omega(\boox{t}\lortild,p))$  
has singularities in the strip $\strip$ for any fixed $p\in\Hyp$ and 
$\lortild\in\Lortild$ in a neighbourhood of the unit. These
singularities are in fact branch points if $s$ is not an integer 
(see Lemma~C.1 in \cite{M02a}). 
However, we have constructed a function $u(p)$ living on the mass
shell which compensates the singularities of the Wigner 
rotation factor. In more detail, our function is given by 
\begin{align} \label{equ} 
  u(p)&:= \Big(\frac{p_0-p_1}{m}\cdot
\frac{{p_0-p_1+m-ip_2}}{p_0-p_1+m+ip_2}\Big)^{s}\;,\quad 
p_0:= ({p_1^2+p_2^2+m^2})^{\half}\,.
\end{align}
(Note that $p_0-p_1$ is strictly positive for all $p\in\Hyp,$ hence the 
argument in brackets lies in the cut complex plane 
$\CC\setminus \RR^-_0.$ The power of $s\in\RR$ is then defined via the
branch of the logarithm on $\CC\setminus \RR^-_0$ with $\ln 1=0.$)   
We then define a map $ c :\Lortild\times\Hyp\rightarrow \CC\setminus
\{0\}$ by 
\begin{align}
 c (\lortild,p)&:= u(p)^{-1}\;e^{is\Omega(\lortild,p)}
\;u(\lor^{-1}\,p) \,.\label{eqDefCoc}
\end{align}
In group theoretical terms, the map 
$c(\cdot,\cdot):\Lortild\times\Hyp\rightarrow \CC\setminus \{0\}$ 
is a cocycle which is equivalent to the Wigner rotation factor.  
To state its analyticity properties, we need some more notation. 
Let $W_1$ be the wedge region  
\begin{equation}\label{eqW1}
W_1 :=\left\{  x\in\mathbb{R}^{3};x^{1}>\left|  x^{0}\right|  \right\},  
\end{equation}
and let the reference direction $\spd_0$ be specified as
$\spd_0=(0,0,-1)$. 
Denote by $\We$ the pair $(W_1,\spdpath_1)$,
where $\spdpath_1$ is the equivalence class of a path in $\Spd$
starting from the reference direction $\spd_0$ and staying within
$W_1$ in the sense of Eq.~\eqref{eqDirInSpc}. If $\spdpath$ is a path
in $\Spd$ ending at a direction $\spd$ contained in $W_1$ in the sense
of Eq.~\eqref{eqDirInSpc}, and $\spdpath$ is equivalent to
$\spdpath_1$ w.r.t\ $W_1$, we write 
\begin{equation} \label{eqDirInSpcPath}
\spdpath\in\We.
\end{equation}
We found the following result. 
\begin{Lem}[\cite{M02a}] \label{CocW1}
Let 
$\lortild$ be an element of $\Lortild$ such that 
$\lortild\act\refspdpath\in\We$ in the sense of 
equation~\eqref{eqDirInSpcPath}. Then for all $p\in\Hyp$ the function   
$$
t\mapsto  c (\boox{t}\lortild,p) 
$$
has an analytic extension into the strip $\RR+i(0,\pi)$.  
This extension satisfies the boundary condition 
\begin{align} \label{eqCociPi}
  c (\boox{i\pi}\lortild,p) & = e^{i\pi s}\,  \overline{c (\lortild,-jp)}\\
& \equiv e^{i\pi s}\, c (\J \lortild \J ,p). \label{eqCocCocj}
\end{align}
\end{Lem}
(The very last equation is not contained in~\cite{M02a}, but follows
directly from the identity~\eqref{eqCocycjj} 
and the fact that the function $u$ 
satisfies $u(-\J p)= \overline{u(p)}$.) 

Let us rewrite this result for the present purpose, namely, the proof of
Lemma~\ref{Transform}. Lemma~\ref{Transform} needs an analyticity
statement for $\lortild$ in a neighbourhood 
of the unit (namely the set $\calUtild_{12}$), whereas the set of 
$\lortild$ satisfying the hypothesis 
of Lemma~\ref{CocW1'} is {\em not} a neighbourhood of the unit (since
$\refspd$ is at the boundary of $W_1$). 
To this end, we fix a Lorentz transformation $\lor_0$ which maps
$\refspd$ into $W_1$, and let  $\lortild_0$ be the (unique) element of
$\Lortild$ over $\lor_0$ such
that $\lortild_0\act\refspdpath\in\We$ in the sense of
equation~\eqref{eqDirInSpcPath}. (For example, a rotation about
$\pi/2$ would do.)   
We then define 
\begin{align} \label{equL0}
u_{\lor_0}(p) & := e^{is\Omega(\lortild_0,p)}
\, u(\lor_0^{-1}p) \equiv u(p)\, c(\lortild_0,p), 
\intertext{ and a corresponding cocycle} 
 c_{\lor_0}(\lortild,p)&:= u_{\lor_0}(p)^{-1}\;e^{is\Omega(\lortild,p)}
\;u_{\lor_0}(\lor^{-1}\,p) \,.\label{eqDefCocL0}
\end{align}
%
\begin{Lem} \label{CocW1'}
i) Let $\lortild$ be an element of $\Lortild$ such that 
$\lortild\lortild_0\act\refspdpath\in\We$ in the sense of 
equation~\eqref{eqDirInSpcPath}.  
Then for all $p\in\Hyp$ the function   
\begin{equation} \label{eqF} 
f(t):=  e^{is\Omega(\boox{t}\lortild,p)}\, 
u_{\lor_0}(\lor^{-1}\Boox{-t}p) 
\end{equation}
has an analytic extension into the strip $\RR+i(0,\pi)$, continuous at
the boundary.  
At $t=i\pi$, this extension has the boundary value 
\begin{align} \label{eqCociPi'} 
f(i\pi) &=  
e^{i\pi s}\, \overline{u_{\lor_0}(- \J  p) \, 
c_{\lor_0}(\lortild,-\J  p)}
\\
& \equiv 
e^{i\pi s}\, e^{is\Omega(\J \lortild \lortild_0\J ,p)}\, 
u\big(\J  (\lor \lor_0)^{-1}\J p\big) . \label{eqCociPi''} 
\end{align}
ii) If $\lor_0$ is the rotation about $\pi/2$,  
then the set of $\lortild$ satisfying the hypothesis of (i) is a 
neighbourhood of the unit. Further, in this case $u_{\lor_0}$ is given
by 
\begin{equation} \label{equPi2}
u_{\lor_0}(p) = e^{is\pihalf} \,  \Big(\frac{p_0-p_2}{m}\cdot
\frac{{p_0-p_2+m+ip_1}}{p_0-p_2+m-ip_1}\Big)^{s}=:u_{\pihalf}(p) \,.
\end{equation}
\end{Lem}
\begin{Proof}
Ad $i)$ 
By definition of the cocycle $c_{\lor_0}$, $f(t)$ coincides with
$u_{\lor_0}(p) \, c_{\lor_0}(\boox{t}\lortild,p)$. 
Since our definitions imply the identity 
\begin{equation} \label{equcL0}
u_{\lor_0}(p) \, c_{\lor_0}(\lortild,p) = 
u(p) \, c(\lortild\lortild_0,p) 
\end{equation}
for all $\lortild\in \Lortild$, we have 
\begin{equation} 
f(t) = u(p) \,   c(\boox{t}\lortild\lortild_0,p) . 
\end{equation}
Lemma~\ref{CocW1} then asserts that for 
$\lortild\lortild_0\act\refspdpath\in \We$, this function is analytic
in the strip $\strip$, and has the boundary value 
\begin{equation} \label{eqFiPi} 
f(i\pi)=  e^{i\pi s}\, u(p) \, \overline{c(\lortild\lortild_0,-\J p)}. 
\end{equation}
Using $u(p)=\overline{u(-\J p)}$ and once again Eq.~\eqref{equcL0}, yields
Eq.~\eqref{eqCociPi'} of the Lemma. 
On the other hand, substituting Eq.~\eqref{eqCocCocj} into
Eq.~\eqref{eqFiPi} and using the defining relation~\eqref{eqDefCoc},
yields Eq.~\eqref{eqCociPi''} of the Lemma.

Ad $ii)$  A rotation $\rot{\pihalf}$ about $\pi/2$ maps $\refspd$ 
into the interior of the wedge $W_1$. Hence the set of $\lortild$ 
satisfying the hypothesis of $(i)$ is a neighbourhood of the unit.
Further, the corresponding $\lortild_0$ is just $\rottild{\pihalf}$,
where $\rottild{\cdot}$ is the lift of the one-parameter group of
rotations to $\Lortild$. Hence $\Omega(\lortild_0,p) = \pi/2$ by 
Eq.~\eqref{eqCocRot}. Together with  
$\rot{\pihalf}^{-1}(p_0,p_1,p_2)= (p_0,p_2,-p_1)$, this implies 
Eq.~\eqref{equPi2}. 
\end{Proof}
\paragraph{Acknowledgements.}
It is a pleasure for me to thank Klaus Fredenhagen for drawing my
attention to the article of Buchholz and Epstein on my search for a
PCT theorem for anyons. 
Further, I gratefully acknowledge financial support by 
FAPEMIG and by the Graduiertenkolleg ``Theoretische  Elementarteilchenphysik'' 
(Hamburg). 

\providecommand{\bysame}{\leavevmode\hbox to3em{\hrulefill}\thinspace}
\providecommand{\MR}{\relax\ifhmode\unskip\space\fi MR }
\providecommand{\MRhref}[2]{%
  \href{http://www.ams.org/mathscinet-getitem?mr=#1}{#2}
}
\providecommand{\href}[2]{#2}

\end{document}